\documentclass[british,english,manuscript,superscriptaddress]{revtex4-1}
\usepackage[T1]{fontenc}
\usepackage[latin9]{inputenc}
\usepackage{babel}
\usepackage{textcomp}
\usepackage{amsmath}
\usepackage{amssymb}
\usepackage{graphicx}
\usepackage[unicode=true,pdfusetitle,
 bookmarks=true,bookmarksnumbered=false,bookmarksopen=false,
 breaklinks=false,pdfborder={0 0 1},backref=false,colorlinks=false]
 {hyperref}
\usepackage{breakurl}

\makeatletter

\newcommand{\lyxmathsym}[1]{\ifmmode\begingroup\def\b@ld{bold}
  \text{\ifx\math@version\b@ld\bfseries\fi#1}\endgroup\else#1\fi}

 \@ifundefined{textcolor}{}
 {%
  \definecolor{BLACK}{gray}{0}
  \definecolor{WHITE}{gray}{1}
  \definecolor{RED}{rgb}{1,0,0}
  \definecolor{GREEN}{rgb}{0,1,0}
  \definecolor{BLUE}{rgb}{0,0,1}
  \definecolor{CYAN}{cmyk}{1,0,0,0}
  \definecolor{MAGENTA}{cmyk}{0,1,0,0}
  \definecolor{YELLOW}{cmyk}{0,0,1,0}
  }

\RequirePackage{colortbl, tabularx}
\@ifundefined{comment}{}
  {
   
  }%

\makeatother

\usepackage{babel}

\makeatother

\begin{document}

\title{Coexistence of ferromagnetism and superconductivity in iron based
pnictides: a time resolved magnetooptical study}

\author{A. Pogrebna}

\affiliation{Complex Matter Dept., Jozef Stefan Institute, Jamova 39, SI-1000
Ljubljana, Slovenia}

\affiliation{Jožef Stefan International Postgraduate School, Jamova 39, SI-1000
Ljubljana, Slovenia}

\author{T. Mertelj}

\email{Correspondence to tomaz.mertelj@ijs.si}

\affiliation{Complex Matter Dept., Jozef Stefan Institute, Jamova 39, SI-1000
Ljubljana, Slovenia}

\author{N.Vuji\v{c}i\'c}

\affiliation{Complex Matter Dept., Jozef Stefan Institute, Jamova 39, SI-1000
Ljubljana, Slovenia}

\affiliation{Institute of Physics, Bijeni\v{c}ka 46, HR-10000 Zagreb, Croatia}

\author{G. Cao}

\affiliation{Department of Physics, Zhejiang University, Hangzhou 310027, People\textquoteright{}s
Republic of China}

\author{Z. A. Xu}

\affiliation{Department of Physics, Zhejiang University, Hangzhou 310027, People\textquoteright{}s
Republic of China}

\author{D. Mihailovic}

\affiliation{Complex Matter Dept., Jozef Stefan Institute, Jamova 39, SI-1000
Ljubljana, Slovenia}

\affiliation{CENN Nanocenter, Jamova 39, SI-1000 Ljubljana, Slovenia}

\date{\today}

\pacs{75.78.Jp, 74.25.Ha, 76.50.+g, 74.70.Xa}
\begin{abstract}
Ferromagnetism and superconductivity are antagonistic phenomena. Their
coexistence implies either a modulated ferromagnetic order parameter
on a lengthscale shorter than the superconducting coherence length
or a weak exchange coupling between the itinerant superconducting
electrons and the localized ordered spins. In some iron based pnictide
superconductors the coexistence of ferromagnetism and superconductivity
has been clearly demonstrated. The nature of the coexistence, however,
remains elusive since no clear understanding of the spin structure
in the superconducting state has been reached and the reports on the
coupling strength are controversial. We show, by a direct optical
pump-probe experiment, that the coupling is weak, since the transfer
of the excess energy from the itinerant electrons to ordered localized
spins is much slower than the electron-phonon relaxation, implying
the coexistence without the short-lengthscale ferromagnetic order
parameter modulation. Remarkably, the polarization analysis of the
coherently excited spin wave response points towards a simple ferromagnetic
ordering of spins with two distinct types of ferromagnetic domains.
\end{abstract}
\maketitle
In the iron-based superconductors family\cite{KamiharaKamihara2006,kamiharaWatanabe2008}
EuFe$_{2}$(As,P)$_{2}$\cite{RenTao2009} and Eu(Fe,Co)$_{2}$As$_{2}$\cite{JiangXing2009}
offer an interesting experimental possibility to study the competition
between the ferromagnetic (FM) and superconducting (SC) order parameters
that can lead to nonuniform magnetic and SC states\cite{AndersonSuhl1959,BuzdinBulaevskii1984,JiangXing2009,BlachowskiRuebenbauer2012}
since the optimal superconducting critical temperature $T\mathrm{_{c}\sim28}$
K\cite{JeevanKasinathan2011} is comparable to the FM Eu$^{2+}$-spin
ordering temperatures $T\mathrm{_{C}\sim18}$ K.\cite{RenZhu2008,RenTao2009} 

The strength and nature of the coupling between the carriers in the
FeAs planes, responsible for superconductivity, and localized Eu$^{2+}$
$f$-orbitals spins, responsible for ferromagnetism, is expected to
influence strongly any possible magnetic as well as SC modulated state.\cite{BuzdinBulaevskii1984}
To enable coexistence of the singlet superconductivity with ferromagnetism
in the case of strong exchange-interaction-dominated coupling the
magnetization modulation period should be short on the lengthscale
below the SC coherence length\cite{AndersonSuhl1959,BuzdinBulaevskii1984},
which is a few\cite{YinZech2009,shanWang2011} tens of nm in 122 iron
based compounds. On the other hand, in the case of weaker long-range
magnetic-dipole dominated coupling a longer lengthscale FM domain
structure can effectively minimize the internal magnetic field enabling
coexistence of the singlet superconductivity and FM state.\cite{FaureBuzdin2005}
Alternatively a spontaneous SC vortex state\cite{BuzdinBulaevskii1984}
might form as proposed recently\cite{NandiJin2014} for EuFe$_{2}$(As,P)$_{2}$.

In the literature opposing claims regarding the coupling between the
carriers in the FeAs planes and localized Eu$^{2+}$ spins exist.
A weak coupling between Fe and Eu magnetic orders was initially suggested
by Xiao \emph{et al.}\cite{xiaoSu2009}, while recently a strong coupling
was suggested from the in-plane magnetoresistance\cite{XiaoSu2012}
and NMR\cite{GuguchiaRoos2011}.

The strength of the coupling between the carriers in the FeAs planes
and localized Eu$^{2+}$ $f$-orbitals spins should be reflected also
in the energy transfer speed between the two subsystems upon photoexcitation.
We therefore systematically investigated the ultrafast transient reflectivity
($\Delta R/R$) dynamics and time resolved magneto-optical Kerr effect
(TR-MOKE) in EuFe$_{2}$(As$_{1-x}$P$_{x}$)$_{2}$ in both, the
undoped spin-density wave (SDW) and doped SC state. In addition to
the relaxation components, that were observed earlier in related non-FM
Ba(Fe,Co)$_{2}$As$_{2}$,\cite{TorchinskyMcIver2011,StojchevskaMertelj2012}
we found another slow-relaxation component associated with Eu$^{2+}$-magnetization
dynamics. The relatively slow 0.1-1 nanosecond-timescale response
of the Eu$^{2+}$ spins to the optical excitation of the FeAs itinerant
carriers \emph{indicates a rather weak coupling} between the two subsystems
suggesting the magnetic-dipole dominated coupling between SC and FM
order parameters.

Moreover, the antiferromagnetic (AFM) Eu$^{2+}$-spin order in the
undoped SDW EuFe$_{2}$As$_{2}$, where the spins are aligned ferromagnetically
in the $ab$ plane with the A-type AFM order of the adjacent Eu$^{2+}$
planes along the $c$ axis, is rather well understood.\cite{RenZhu2008,xiaoSu2009}
Contrary, no coherent picture of Eu$^{2+}$-spin ordering upon P or
Co doping exists. In addition to the proposal of a SC induced helimagnetic
ordering\cite{JiangXing2009} in Eu(Fe,Co)$_{2}$As$_{2}$ a canted
AFM was proposed by Zapf \emph{et al.} \cite{ZapfWu2011} in superconducting
EuFe$_{2}$(As$_{1-x}$P$_{x}$)$_{2}$, while a pure FM ordering\cite{NandiJin2014}
at $x=0.15$ coexisting with superconductivity was reported by Nandi
\emph{et al.} \cite{NandiJin2014}. A spin-glass state over the all
P doping range was also suggested by Zapf \emph{et al.} \cite{ZapfJeevan2013}
recently. 

The observed time-resolved magnetooptical transients in the presence
of an in-plane magnetic field reveal an additional coherent magnon
response in the superconducting sample. The polarization dependence
of the coherent magnon oscillations \emph{points towards a FM domain
state} consistent with results of Nandi \emph{et al.} \cite{NandiJin2014}\emph{.}

\section*{Results}

\begin{figure}[tbh]
\begin{centering}
\includegraphics[bb=0bp 50bp 595bp 760bp,clip,width=0.8\columnwidth]{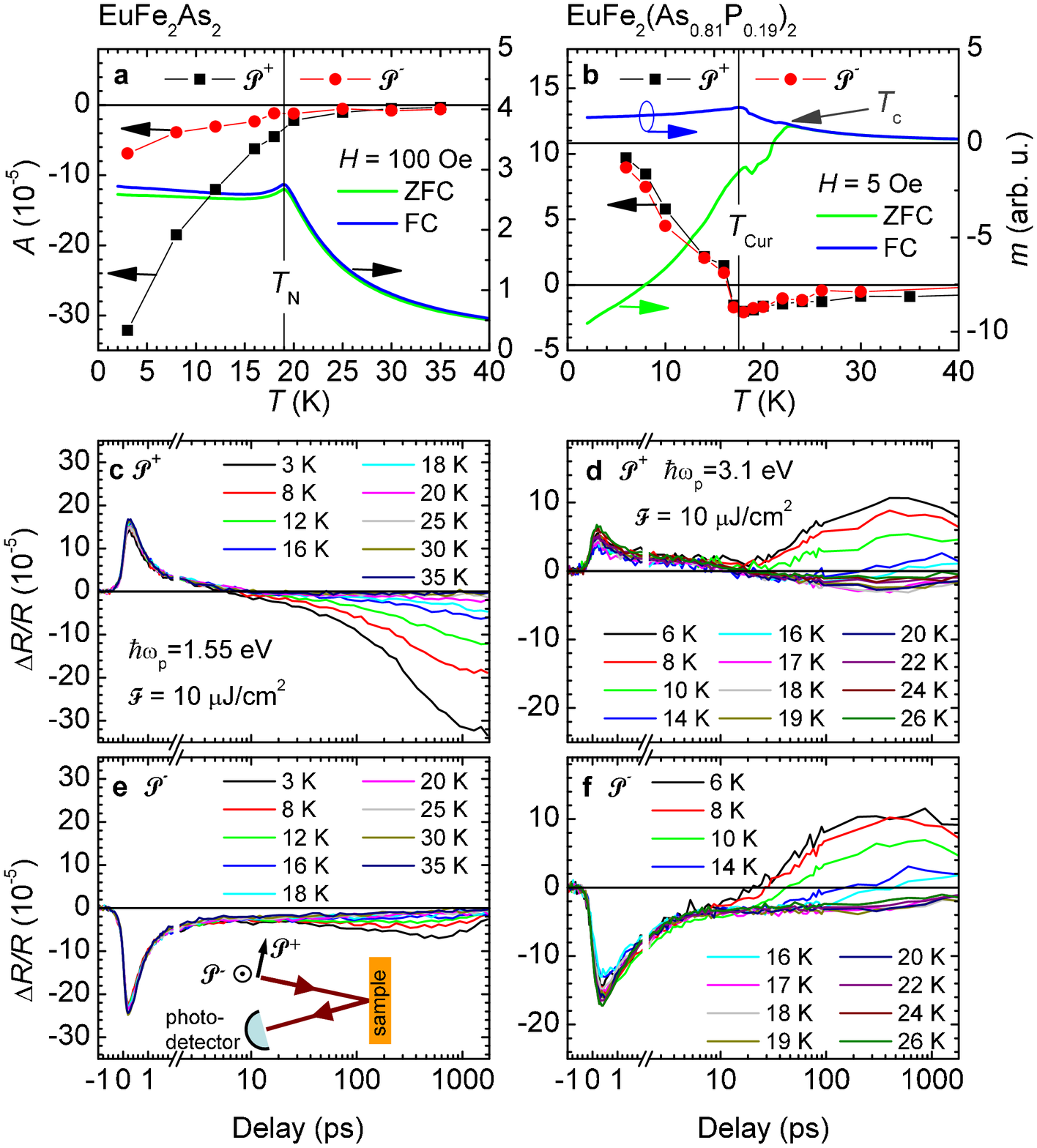} 
\par\end{centering}

\caption{Temperature dependence of the $ab$-plane transient reflectivity.
The amplitude of the photoinduced reflectivity transients at long
delays as a function of temperature in EuFe$_{2}$As$_{2}$, (\textbf{a}),
and EuFe$_{2}$(As$_{0.81}$P$_{0.19}$)$_{2}$, (\textbf{b}), compared
to the magnetic moment along the $c$-axis . $\mathcal{P}^{+}$ and
$\mathcal{P}^{-}$ correspond to two orthogonal in-plane probe-photon
polarizations while ZFC and FC correspond to cooling in the presence
and absence of magnetic field, respectively. Photoinduced reflectivity
transients at low-$T$ in EuFe$_{2}$As$_{2}$ (\textbf{c}), (\textbf{e})
and EuFe$_{2}$(As$_{0.81}$P$_{0.19}$)$_{2}$ (\textbf{d}), (\textbf{f})
for the two probe-photon polarizations. Inset to (\textbf{e}) represents
a schematic of the probe beam configuration.}

\label{fig:DRvsT} 
\end{figure}

\textbf{Temperature dependence of photoinduced reflectivity.} In Fig.
\ref{fig:DRvsT} c)-f) we show temperature dependence of the transient
reflectivity ($\Delta R/R$) measured with the probe pulses polarized
in the $ab$-plane in undoped nonsuperconducting EuFe$_{2}$As$_{2}$
(Eu-122) and doped superconducting EuFe$_{2}$(As$_{0.81}$P$_{0.19}$)$_{2}$
(EuP-122). The transient reflectivity is anisotropic in the $ab$-plane,
consistent with the orthorhombic crystal structure. We indicate the
two orthogonal polarizations $\mathcal{P}^{+}$ and $\mathcal{P}^{-}$
according to the sign of the subpicosecond transient reflectivity.
In addition to the anisortopic fast component associated with the
SDW order discussed elsewhere\cite{PogrebnaVujicic2014} we observe
in both samples, concurrently with emergence of the Eu$^{2+}$-spin
ordering,\cite{xiaoSu2009,ZapfWu2011} appearance of another much
slower relaxation component {[}see Fig. \ref{fig:DRvsT} a) and b){]}
with a risetime of $\sim1$ ns in Eu-122 and $\sim100$ ps in EuP-122
(at $T=1.5$ K) and the decay time beyond the experimental delay range.
In the vicinity of the Eu$^{2+}$ magnetic ordering temperatures a
marked increase of the risetime is observed in both samples. In Eu-122
the slow component is rather anisotropic, while in EuP-122 it appears
almost isotropic. 

\begin{figure}
\includegraphics[bb=0bp 320bp 570bp 842bp,clip,width=0.8\columnwidth]{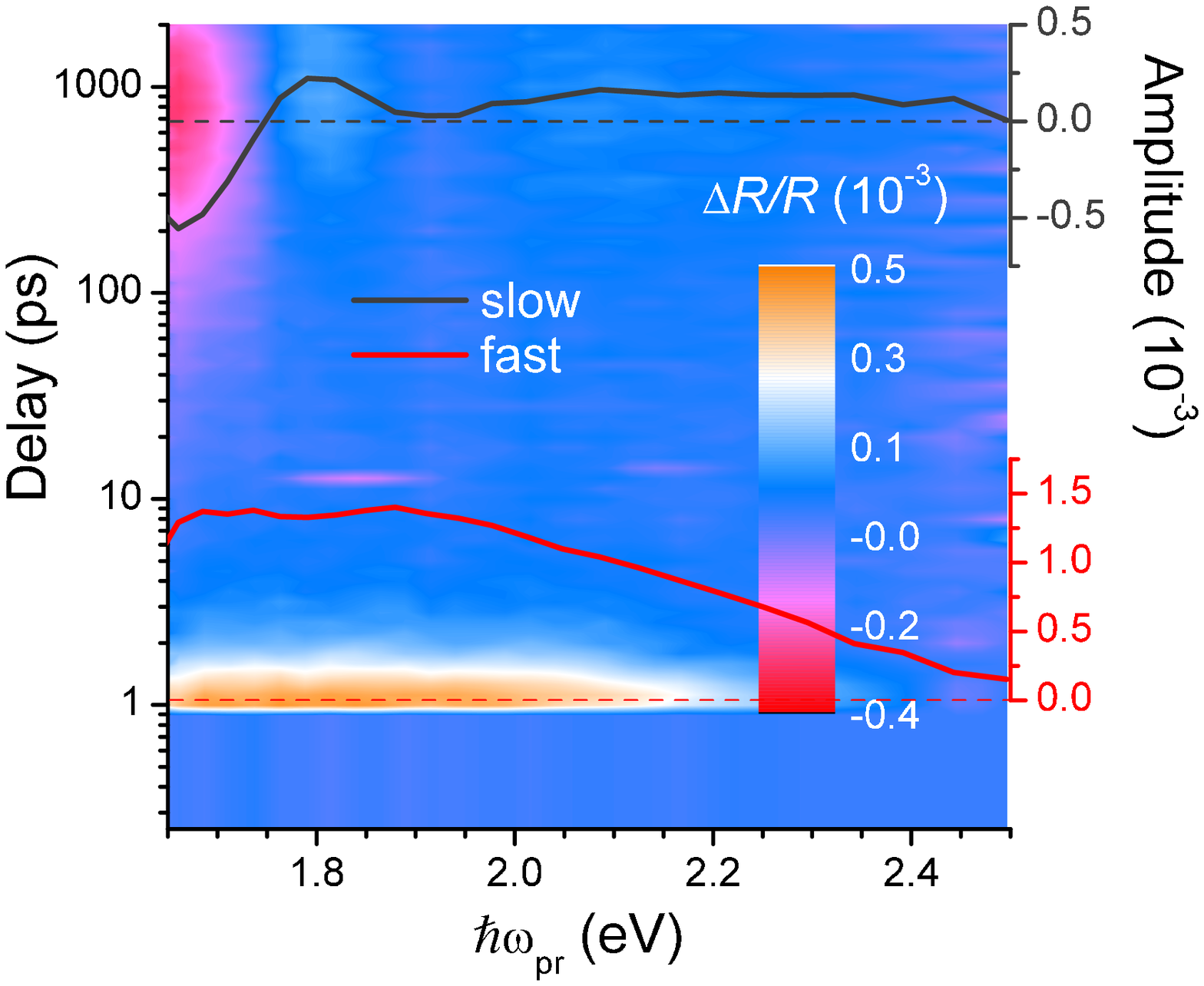}

\caption{Photoinduced reflectivity transients at low-$T$ in EuFe$_{2}$As$_{2}$
as a function of probe photon energy for the $\mathcal{P}^{+}$ polarization.
The spectral dependencies of the fast and slow response amplitude
are shown as red and \foreignlanguage{british}{dark-grey} lines, respectively.}
\label{fig:figDispersionEFA}
\end{figure}

The probe-photon-energy dependence of the transients in Eu-122 is
shown in Fig. \ref{fig:figDispersionEFA}. The dispersion of the fast
component\cite{PogrebnaVujicic2014} is much broader than that of
the slow one, which shows a relatively narrow resonance around $\sim$1.7
eV.

\begin{figure}[tbh]
\begin{centering}
\includegraphics[bb=0bp 70bp 595bp 842bp,angle=-90,width=1.1\columnwidth]{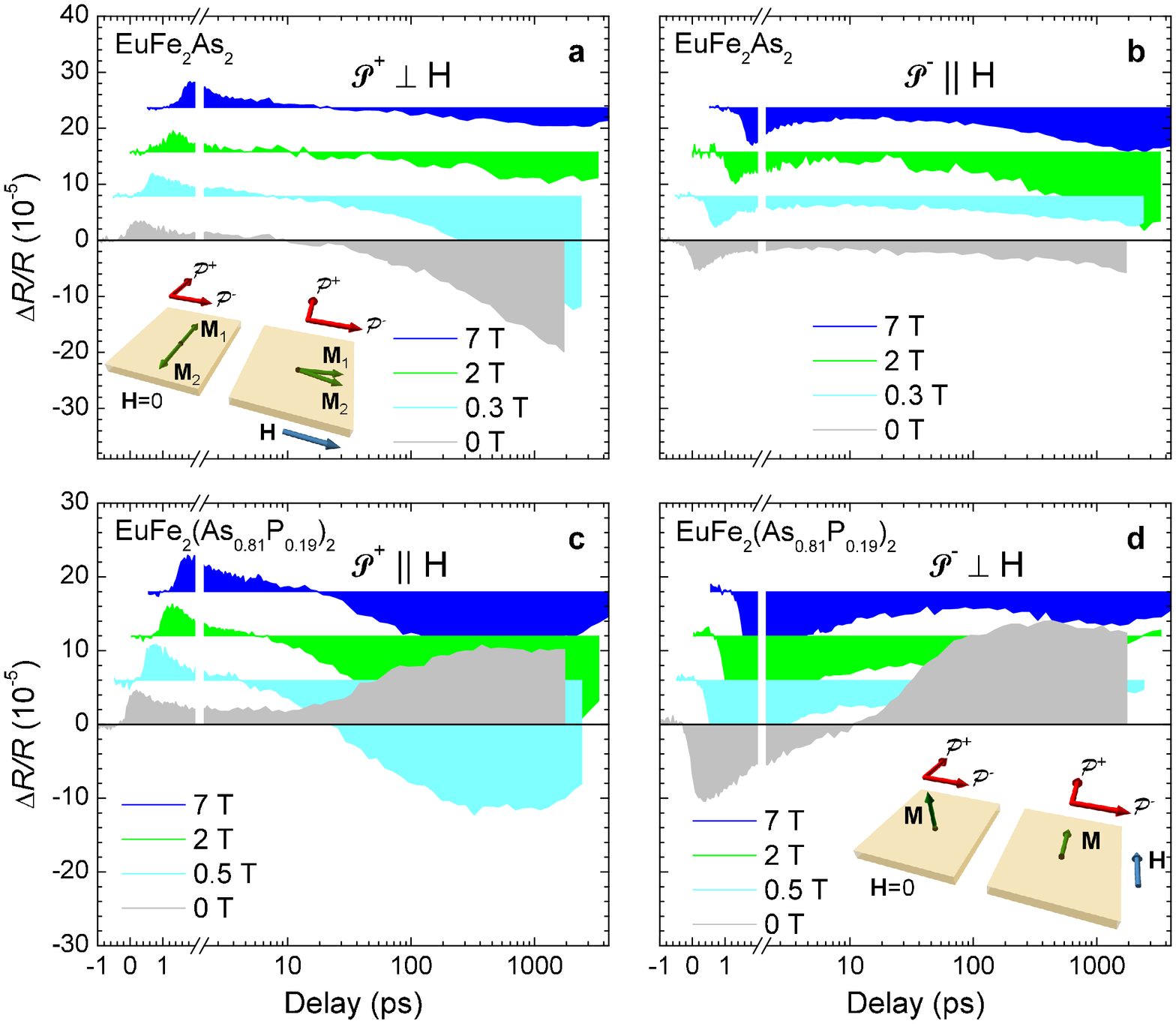} 
\par\end{centering}

\caption{In-plane magnetic field dependence of the transient reflectivity.
(\textbf{a}), (\textbf{b}) The reflectivity transients in EuFe$_{2}$As$_{2}$
with the magnetic field field paralel to the $\mathcal{P}^{-}$ polarization.
(\textbf{c}), (\textbf{d}) The reflectivity transients in EuFe$_{2}$(As$_{0.81}$P$_{0.19}$)$_{2}$
with the magnetic field field paralel to the $\mathcal{P}^{+}$ polarization.
All transient were measured at $T=2$ K, $\mathcal{F}\sim3$ $\mu$J/cm$^{2}$
and 1.55-eV pump-photon energy. Insets show shematically magnetization
reorientation in magnetic field.}

\label{fig:DRvsB} 
\end{figure}

\textbf{Metamagnetic transitions.} Upon application of magnetic field
lying in the $ab$-plane the Eu$^{2+}$ AFM order in Eu-122 is destroyed
above $\mu_{0}H\sim0.8$ T in favor of an in-plane field-aligned FM
state.\cite{JiangLuo2009,XiaoSu2010,GuguchiaBosma2011} In EuP-122
a similar field-induced spin reorientation from the out-of-plane FM
into the in-plane field-aligned FM state was observed around $\mu_{0}H\sim0.6$
T.\cite{NandiJin2014} These metamagnetic transitions have remarkable
influence on the transient reflectivity as shown in Fig. \ref{fig:DRvsB}.
While the fast picosecond response associated with the SDW state\cite{PogrebnaVujicic2014}
shows virtually no dependence on the magnetic field, the slow response
shows a marked change in the field-induced FM state\cite{JiangLuo2009,XiaoSu2010,GuguchiaBosma2011,NandiJin2014}. 

In undoped Eu-122 the $\mathcal{P}^{+}$-polarization slow response
is suppressed above the metamagnetic transition {[}Fig. \ref{fig:DRvsB}
(a){]} and is magnetic-field independent above 2 T. Concurrently,
for the $\mathcal{P}^{-}$ polarization, which is parallel to the
magnetic field, {[}Fig. \ref{fig:DRvsB} (b){]} the slow response
is first enhanced at low magnetic field above the transition, resembling
a rotation of the anisotropy by $\pi/2$, and then slightly suppressed
upon increasing the field to 7 T.

In EuP-122 the initially positive rather isotropic slow response {[}Fig.
\ref{fig:DRvsB} (c), (d){]} switches to a negative anisotropic one
along the $\mathcal{P}^{+}$ polarization, parallel to the magnetic
field. Similar to Eu-122 the slow response is slightly suppressed
at the highest field with a faster relaxation. 

\begin{figure}[tbh]
\begin{centering}
\includegraphics[bb=0bp 250bp 842bp 1191bp,clip,width=0.9\columnwidth]{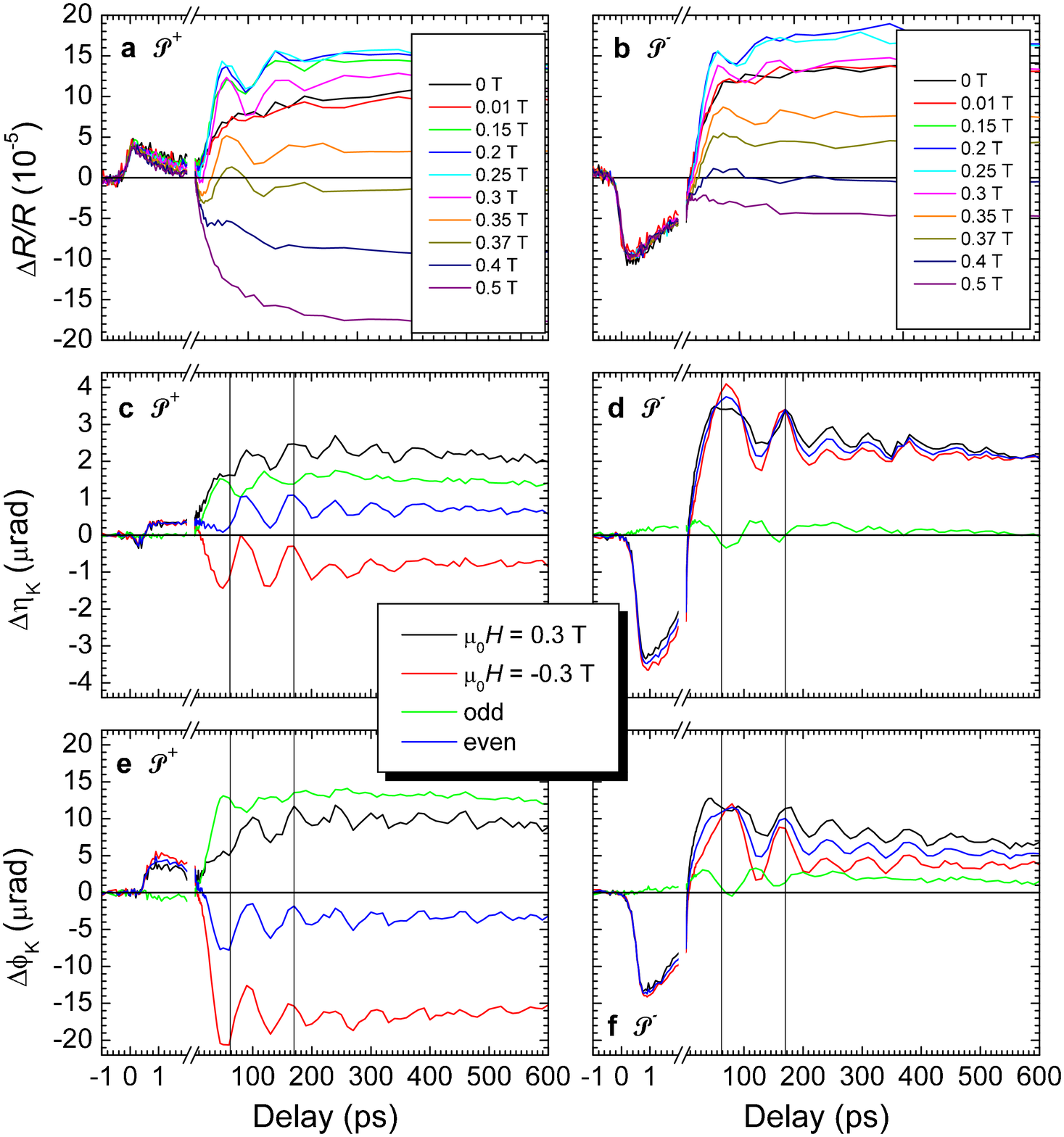} 
\par\end{centering}

\caption{(\textbf{a}), (\textbf{b}) The reflectivity transients in EuFe$_{2}$(As$_{0.81}$P$_{0.19}$)$_{2}$
in low magnetic fields at $T=2$ K, $\mathcal{F}=3$ $\mu$J/cm$^{2}$
and 1.55-eV pump photon energy. Transient Kerr ellipticity, (\textbf{c}),
(\textbf{d}), and rotation, (\textbf{e}), (\textbf{f}), upon reversal
of the magnetic field at $T=1.5$ K and $\mathcal{F}=10$ $\mu$J/cm$^{2}$.
Odd and even part of the responses correspond to the difference an
the sum of the responses measured at different signs of the magnetic
field, respectively. }

\label{fig:DR-MOKE-vsB-osc} 
\end{figure}

\begin{figure}[tbh]
\begin{centering}
\includegraphics[angle=-90,width=0.98\columnwidth]{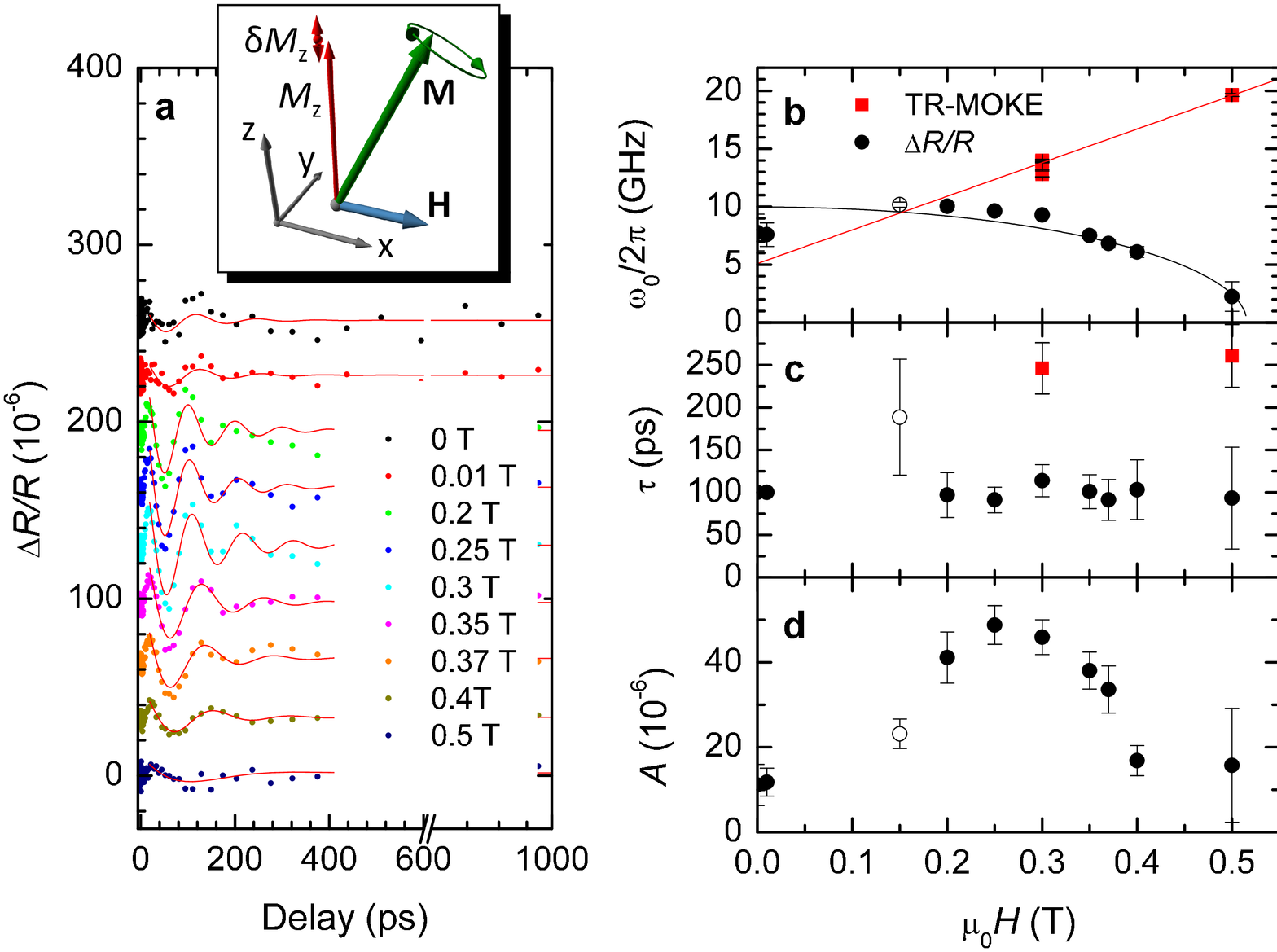} 
\par\end{centering}

\caption{(\textbf{a}) The oscillatory part of the isotropic $\Delta R/R$ component
in EuFe$_{2}$(As$_{0.81}$P$_{0.19}$)$_{2}$ at low magnetic fields,
$\mathcal{F}=3$ $\mu$J/cm$^{2}$ and 1.5-eV pump photon energy.
Thin lines represent the damped oscillator fits discussed in text.
The frequency (b), decay time (c) and amplitude (\textbf{e}) of the
oscillations as functions of the magnetic field. The points (open
symbols) at $B=0.15$ T were obtained from the $\mathcal{P}^{-}$
polarization fit due to the lack of data at the $\mathcal{P}^{+}$
polarization. The red squares were obtained from TR-MOKE fits. The
lines in (\textbf{b}) are uniaxial ferromagnet\cite{turov1965physical}
fits discussed in text. The inset to (\textbf{a}) shematically shows
magnetization precession in small magnetic fields with corresponding
projections onto the $z$-axis.}

\label{fig:figDROscfitB} 
\end{figure}

\textbf{Coherent spin waves.} In EuP-122 at low magnetic fields below
$\sim0.5$ T additional damped oscillations appear on top of the slow
relaxation in $\Delta R/R$ {[}see Fig. \ref{fig:DR-MOKE-vsB-osc}
(a), (b){]}. These oscillations appear rather isotropic. The amplitude
of the oscillations, shown in Fig. \ref{fig:figDROscfitB} (d), is
strongly peaked around $\sim0.25$ T and vanishes at 0.5 T. The frequency
of the oscillations, as determined by a damped oscillator fit shown
in Fig. \ref{fig:figDROscfitB} (a), is $H$ independent at low fields
and starts to decrease with increasing field above $\mu_{0}H\sim0.3$
T. The damping, on the other hand, is magnetic-field independent at
$\tau^{^{-1}}\sim10$ GHz. 

Another oscillation with a higher frequency ($\sim14.5$ GHz at 0.3
T) is revealed by the transient magnetooptical Kerr effect (TR-MOKE)
shown in Fig. \ref{fig:DR-MOKE-vsB-osc} (c)-(f). The oscillatory
part of the transient rotation and ellipticity is polarization independent
and almost even with respect to the reversal of magnetic field.

\section*{Discussion}

Eu$^{2+}$ ions have {[}Xe{]}$4f^{7}$$6s^{2}$ ($^{8}$S$_{7/2}$)
electronic configuration. The lowest excited states of a free Eu$^{2+}$
ion are $\sim3.5$ eV above the ground state.\cite{Dorenbos2003}
In oxides, however, this splitting can be reduced down to $\sim1$
eV.\cite{Feinleib1969,Dorenbos2003} In Eu-122 the position of $f$-derived
states was calculated to be $\sim2$ eV below the Fermi energy,\cite{LiZhu2012}
close to the observed Eu$^{2+}$-spin ordering related slow-component
resonance around 1.7 eV {[}see Fig. \ref{fig:figDispersionEFA} (b){]}.
It is therefore plausible that the coupling of the Eu$^{2+}$ magnetism
to the dielectric constant at the probe photon energy of 1.55 eV is
through the resonant magneto-optical Cotton-Mouton effect with the
location of the Eu$^{2+}$-4$f$ states $\sim1.7$ eV below the Fermi
level.

On the other hand, a large magnetostriction is indicated from the
realignment of the crystal twin domain structure in magnetic field,\cite{XiaoSu2010}
suggesting a possibility of the indirect contribution to the optical
dielectric function through the magnetoelastic effect. The rather
narrow probe-photon-energy resonance of the slow component does not
support this mechanism. 

We should also note that the realignment of the twin domain structure\cite{XiaoSu2010}
was not observed in our experiment, since the anisotropy of the fast
component, which is associated with the structural twin domains,\cite{StojchevskaMertelj2012}
shows no dependence on magnetic field in both samples (see Fig. \ref{fig:DRvsB})
up to $\mu_{0}H=7$ T. Moreover, the realignment of the twin domain
structure observed in Ref. {[}\onlinecite{XiaoSu2010}{]} might be
related to the Fe spin ordering as indicated by observation of a partial
magnetic field detwinning also in non-ferromagnetic Ba(Fe$_{1\lyxmathsym{\textminus}x}$Co$_{x}$)$_{2}$As$_{2}$.\cite{ChuAnalytis2010prb}

The strong in-plane anisotropy of the slow component in Eu-122 indicates
that the response corresponds to the dynamics of the in-plane component
of the sublattice Eu$^{2+}$ magnetizations. The presence of qualitatively
same response in the in-plane field-aligned FM state suggests that
the observed slow dynamics is not the dynamics of the AFM order parameter,
but rather the dynamics of the individual AFM sublattice magnetizations.
The response can therefore be associated with a decrease of the Eu$^{2+}$
magnetization upon photoexcitation in both, the zero-field AFM and
the field-induced in-plane FM state.

To understand the change of the anisotropy between weak and strong
magnetic fields in EuP-122 let us look at the symmetric part of the
in-plane dielectric tensor components $\epsilon_{ii}$. Within the
orthorhombic point symmetry $\epsilon_{ii}$ can be expanded in terms
of magnetization to the lowest order as:
\begin{equation}
\epsilon_{ii}=\epsilon_{0,ii}+a_{iizz}M_{z}^{2}+a_{iixx}M_{x}^{2}+a_{iiyy}M_{y}^{2},\label{eq:epssym}
\end{equation}
with $i\in\{x,y\}$. Here $\mathbf{M}$ would correspond to the Eu$^{2+}$
sublattice magnetization in the case of a canted AFM ordering, or
the total Eu$^{2+}$ magnetization in the case of FM ordering. In
EuP-122 in low magnetic fields $\mathbf{M}$ is predominantly oriented
along the $c$-axis\cite{NowikFelner2011jp,NandiJin2014} leading
to the nearly isotropic response, since $a_{xxzz}\sim a_{yyzz}$ due
to the small orthorhombic lattice distortion. In the field-induced
FM state and the zero-field AFM state of Eu-122 $\mathbf{M}$ lies
in the $ab$-plane leading to an anisotropic response since it is
quite unlikely that $a_{iiii}\sim a_{jjii}$, with $i\neq j$.

The photoinduced Eu$^{2+}$ demagnetization is therefore slow, on
a nanosecond timescale in Eu-122 and a $\sim100$ ps timescale in
EuP-122. It can not be due to a direct emission of incoherent Eu$^{2+}$
magnons by the eV-energy photoexcited Fe-$d$-bands electron-hole
pairs since it has been shown, that in the case of iron-based pnictides
in the SDW state the Fe-$d$-bands quasiparticle relaxation occurs
on a picosecond timescale\cite{StojchevskaKusar2010,RettigCortes2013,PogrebnaVujicic2014}
and goes through emission of Fe-$d$-spin magnons\cite{PogrebnaVujicic2014}
followed by relaxation to phonons. It can therefore be assumed that
the Fe-$d$-bands quasiparticle and lattice degrees of freedom are
fully thermalized beyond $\sim10$ ps when the slow component starts
to emerge. This suggests that\emph{ the energy transfer from the excited
quasiparticles in the Fe-$d$ bands to the Eu$^{2+}$ magnons is rather
inefficient}. The incoherent Eu$^{2+}$ magnons are therefore excited
indirectly via the spin-lattice coupling only after the initial excitation
energy was thermally distributed between the Fe-$d$-bands quasiparticles
and phonons. \emph{The Eu$^{2+}$ spins therefore appear only weakly
coupled to the Fe-$d$-bands quasiparticles with the coupling increasing
with the P doping. }The rather large in-plane magnetoresistance observed
by Xiao \emph{et al.} {[}\onlinecite{XiaoSu2012}{]} in Eu-122 can
therefore be attributed to slow magnetostriction effects modifying
the lattice twin domain structure.

The light penetration depth at the probe-photon energy of $\hbar\omega_{\mathrm{pr}}=1.55$
eV is $\sim27$ nm,\cite{WuBarisic2009} while the beam diameters
are in a 100 $\mu$m range. Irrespective of the excitation mechanism,
which can be either nonthermal impulsive\cite{HansteenKimel2005}
inverse Cotton-Mouton effect or thermal displacive non-Raman\cite{KalashnikovaKimel2008}
like, it can be assumed that the relevant wavevectors are $q\lesssim1/30$
nm$^{-1}$ and dominantly a uniform coherent magnetization precession
is excited and detected. (In the case of helical magnetic order with
the propagation vector ${\bf q}_{0}$, spin waves at ${\bf q}=\pm m{\bf q}_{0}$,
$m\in\mathbb{Z}$, also need to be considered.\cite{CooperElliot1963})

The low frequency mode observed in the transient reflectivity response
softens with increasing temperature and vanishes in the field induced
in-plane FM state so it can definitely be assigned to a magnetic mode.
The high frequency mode has also a magnetic origin since it appears
in the TR-MOKE configuration only. 

Analyzing contributions of the magnetization displacements, $\delta M_{i}$,
to the symmetric part of the optical response it follows from (\ref{eq:epssym}),

\begin{equation}
\delta\epsilon_{ii}=2a_{iizz}M_{z}\delta M_{z}+2a_{iixx}M_{x}\delta M_{x}+2a_{iiyy}M_{y}\delta M_{y}.\label{eq:deppssym}
\end{equation}
The low-frequency mode is very strong in $\Delta R/R$ and rather
isotropic in the $ab$-plane indicating that is either associated
with the out of-plane terms (i) $2a_{iizz}M_{z}\delta M_{z}$ or (ii)
both, $M_{x}$ and $M_{y}$, are finite such as in the case of a helimagnetic
ordering. In the latter case the local magnetization needs to be considered
since the average of the terms <$M_{i}\delta M_{i}$>, $i\in\left\{ x,y\right\} $,
over Eu$^{2+}$ planes is finite despite <$M_{i}$>$=0$. Concurrently,
it is weak in the TR-MOKE configuration, which is sensitive to $\delta M_{z}$.
Since $\delta M_{z}\not=0$ for both (i) and (ii) (see Supplemental
information for case (ii)) this indicates that the measured volume
is composed from the ``up'' and ``down'' magnetic domains magnetized
along the $c$-axis. The sign of $\delta M_{z}$ varies in different
magnetic domains leading to a vanishing TR-MOKE response averaged
over many magnetic domains, while the sign of $M_{z}\delta M_{z}$
does not depend on the domain orientation and averages to a finite
value.

For the high-frequency mode observed in the TR-MOKE configuration,
on the other hand, the averaged $\delta M_{z}$ is finite while the
averaged $M_{z}\delta M_{z}$ is rather small in comparison to the
low frequency mode. This indicates that in addition to the $c$-axis
magnetized domains in-plane magnetized regions exist with $M_{z}\sim0$.
The in-plane magnetization leads to an out of plane magnetization
displacement with $\delta M_{z}\neq0$ and $M_{z}\delta M_{z}\sim0$,
consistent with the observed magnetic field dependence of the mode
frequency. The invariance of the oscillatory TR-MOKE response {[}see.
Fig. \ref{fig:DR-MOKE-vsB-osc} (c)-(f){]} with respect to the inversion
of the magnetic field is also consistent with the in-plane magnetization
orientation. 

A fit of the frequency magnetic-field dependence {[}see Fig. \ref{fig:figDROscfitB}
(b){]} using the standard uniaxial ferromagnet formula for a parallel
magnetic field\cite{turov1965physical} ignoring demagnetization factors,
$\omega=\gamma_{ab}(H_{\mathrm{ab}}+H)$, yields $\mu_{0}H_{\mathrm{ab}}=0.3$
T and $\gamma_{ab}/\mu_{0}=182$ GHz/T. The obtained gyromagnetic
ratio $g_{ab}=2.06$ is consistent with $^{8}$S$_{7/2}$ state of
Eu$^{2+}$ ions. The absence of demagnetization factors suggests that
the response does not originate from the domain walls between the
$c$-axis oriented domains but rather from planar shaped domains.
Due to surface sensitivity ($\sim30$ nm) of the optical probe these
are very likely surface domains, however, the bulk nature of these
domains can not be entirely excluded.

The observed \foreignlanguage{british}{behaviour} is compatible with
the simple ferromagnetic order (within the domains) proposed by Nandi
\emph{et al.}\cite{NandiJin2014}. In the absence of the in-plane
magnetic field the static magnetization in the $c$-axis domains is
along the $c$-axis and $\delta M_{z}=0$, consistent with the vanishing
amplitude of the low frequency mode near the zero field. Upon application
of the in-plane magnetic field the magnetization is tilted away from
the $c$-axis (see inset to Fig. \ref{fig:figDROscfitB}) leading
to a finite $\delta M_{z}$ and the observed decrease of the mode
frequency.\cite{turov1965physical} The decrease of the transient-reflectivity
amplitude, when approaching to the metamagnetic transition, can be
associated with the vanishing $M_{z}$. 

A fit of the frequency magnetic-field dependence using the standard
uniaxial ferromagnet formula for the perpendicular magnetic field\cite{turov1965physical}
ignoring demagnetization factors, $\omega=\gamma_{c}\sqrt{H_{\mathrm{c}}^{2}-H^{2}}$,
results in $\mu_{0}H_{c}=0.52$ T and $\gamma_{c}/\mu_{0}=119$ GHz/T.
The small value of $\gamma_{c}$ leading to a small gyromagnetic ratio
($g_{c}=1.35$) can be attributed to the ignored unknown demagnetization
factors of the $c$-axis magnetized domains. Moreover, it suggests
that the $c$-axis magnetized domains have a flat shape with the normal
perpendicular to the $c$ axis.

On the other hand, the presence of two distinct modes and the magnetic
field dependence of the mode frequencies {[}see Fig. \ref{fig:figDROscfitB}
(b){]} resembles the standard uniaxial AFM cases with the magnetic
field perpendicular to the easy/hard axis\cite{turov1965physical,SubkhangulovHenriques2014}
indicating a possible canted AFM\cite{ZapfWu2011} (CAFM) order. The
polarization dependence of the modes is, however,\emph{ not compatible}
with the CAFM picture since both, the quasi-AFM mode\cite{IidaSatoh2011}
and the quasi-FM mode, contribute to $\delta M_{z}$ (see Supplemental)
and should, contrary to the observations, contribute concurrently
to the transient reflectivity and the TR-MOKE with identical relative
amplitudes. 

In the case of the conical helimagnetic ordering\cite{JiangXing2009,NowikFelner2011,NowikFelner2011jp}
the in-plane isotropy naturally appears for certain modes (see Supplemental).
However, since, as in the case of the CAFM state, contributions of
more than one magnetic mode to $\delta M_{z}$ are expected, our data
do not support the conical helimagnetic ordering.

In conclusion, our data point towards the simple FM Eu$^{2+}$-spin
order in superconducting EuFe$_{2}$(As,P)$_{2}$ proposed by Nandi
\emph{et al.} {[}\onlinecite{NandiJin2014}{]}. The observed weak
coupling between the FeAs-plane quasiparticles and Eu$^{2+}$ spins
indicates a weak magnetic-dipole dominated coupling between the SC
and FM order parameters. This indicates that the coexistence of the
singlet superconductivity with ferromagnetism in EuFe$_{2}$(As$_{1-x}$P$_{x}$)$_{2}$
is possible without necessity of the magnetic structure modulation
on the lengthscale shorter than the SC coherence-length. The presence
of the FM domain structure on longer lengthscales, which is inferred
from the coherent-spin-wave response, might additionally contribute
to stability of the coexisting state.

\section*{Methods}

\textbf{Sample preparation.} Single crystals of EuFe$_{2}$(As$_{1-x}$P$_{x}$)$_{2}$
were grown by a flux method, similar to a previous reports\cite{JiaoTao2011,PogrebnaVujicic2014}
The out-of-plane magnetic susceptibilities shown in Fig. \ref{fig:DRvsT}
are consistent with previous results. \cite{JiangLuo2009,ZapfJeevan2013}
From the susceptibility we infer Eu$^{2+}$ spin ordering temperatures
$T\mathrm{_{N}}=19$ K and $T_{\mathrm{Cur}}=17.6$ K in EuFe$_{2}$As$_{2}$
(Eu-122) and EuFe$_{2}$(As$_{0.81}$P$_{0.19}$)$_{2}$ (EuP-122),
respectively. EuP-122 also shows the onset of superconductivity at
$T\mathrm{_{c}=}22.7$ K.

\textbf{Optical measurements.} Measurements of the photoinduced transient
reflectivity, $\Delta R/R$, from $ab$ facets of freshly cleaved
samples at nearly normal incidence were performed using a standard
pump-probe technique, with 50 fs optical pulses from a 250-kHz Ti:Al$_{2}$O$_{3}$
regenerative amplifier seeded with an Ti:Al$_{2}$O$_{3}$ oscillator.\cite{StojchevskaMertelj2012}
We used the pump photons with both, the laser fundamental ($\hbar\omega_{\mathrm{P}}=1.55$
eV) and the doubled ($\hbar\omega_{\mathrm{P}}=3.1$ eV) photon energy,
and the probe photons with the laser fundamental ($\hbar\omega_{\mathrm{pr}}=1.55$
eV) photon energy.

\textbf{Magnetooptical measurements.} Transient Kerr rotation, $\Delta\phi_{\mathrm{K}}$,
was also measured on $ab$ facets of freshly cleaved samples at nearly
normal incidence by means of a balanced detector scheme using a Wollaston
prism and a standard homodyne modulation technique in a 7-T split-coil
optical superconducting magnet. To measure the transient Kerr ellipticity,
$\Delta\eta_{\mathrm{K}}$, a $\lambda/4$-waveplate was inserted
in front of the Wollaston prism. In order to minimize the pollution
of the Kerr signals with the photoinduced reflectivity signal the
detector was carefully balanced prior to each scan.

\begin{acknowledgments}
Work at Jozef Stefan Institute was supported by ARRS (Grant No. P1-0040).
Work done in Zhejiang University was supported by NSFC (Grant No.11190023)
\end{acknowledgments}

\section*{Author contributions}

T. M. conceived the idea and experiment, A. P. performed optical measurements,
N. V. built broadband optical setup, G. C. and Z.-A. X. grew and characterized
single crystals, T. M. and A. P. analyzed the data, T. M., A. P. and
D. M. wrote the paper.

\section*{Additional information}

\textbf{Competing financial interests:} The authors declare no competing
financial interests.
\end{document}